# Georg(e) Placzek: a bibliometric study of his scientific production and its impact

*Manuel Cardona and Werner Marx*
*Max Planck Institute for Solid State Research, D-70569 Stuttgart (Germany)*

**Abstract:** The availability of a number of databases, in particular the *Science Citation Index* (SCI), have encouraged the development and use of bibliometric techniques to analyze and evaluate the production and impact of scientists. To avoid pitfalls and their sometimes serious consequences, however, considerable experience with the method is needed. The case of George Placzek appears as an excellent one to illustrate the procedure and its problems. Placzek's work covered a broad range of topics, including optical and neutron spectroscopy, neutron diffusion, nuclear reactions, and nuclear energy. He worked in a large number of places with some of the most outstanding collaborators and also as sole author during his short professional life. His publications appeared in regular, so-called source journals, in books, lecture notes and also internal reports which were classified till several years after the end of the war. In this article we analyze Placzek's work and its impact with the aim of illustrating the power and virtues of bibliometric techniques and their pitfalls.

## 1. Introduction

The term *bibliometry* is usually applied to the quantitative investigation of the number of publications of individuals, institutions and/or disciplines and their impact as measured by the number of citations they received. The origin of modern bibliometry is related to the foundation in 1954, by E. Garfield, of the company *Eugene Garfield Associates*, one year before Placzek's untimely death in 1955. In 1960 the company's name was changed to *Institute of Scientific Information* (ISI). Its main product at the time was *Current Contents*, a booklet containing the table of contents of journals selected to be relevant to the progress of science.[1] In 1964 the ISI launched the *Science Citations Index* (SCI), covering at that time 600 scientific journals. In 1988 the printed SCI was complemented by an electronic version and in 1992 the ISI, based in Philadelphia, was bought by the *Thomson Corporation.* The next important step was the launching of the *Web of Science* (WoS) in 1997, available through the internet and covering about 7000 so-called source journals in all fields of scholarship. For a detailed discussion of bibliometric techniques and citation analysis see Moed [2].

The *Web of Science* is based on the ISI citation indexes, in particular on the SCI. The WoS has probably become the most versatile and user friendly citation analysis tool and it is often institutionally available to researchers of many organizations. Its competent use, however, requires some experience and awareness of possibilities and pitfalls, such as interference with namesakes and access to citations of books (usually not source items) provided the citations appear in source items. The latter can be performed using the *Cited Reference Search* mode (see below).

In spite of his short scientific life (about 25 years) Placzek is an excellent subject for learning the pitfalls and tricks of citation analysis. He only published a few articles in source journals (about 30) and many of his citations belong to books, book articles and classified reports made available after the war or even after his death. His name is not very common (we have only found two namesakes cited both after his death).[3,4,5] He worked and studied in a wide range of European cities - Vienna, Prague, Utrecht (1929), Leipzig (1930), Rome (1931-1932), Copenhagen (1932-1938), Paris, Kharkov (1936) - and, after his forced emigration in 1938, in Canada (Chalk River) and the United States (Los Alamos, Schenectady, Ithaca, Princeton, and most likely others) and even Jerusalem (1935). He entertained close connections with some of the leading physicists of the times, as will be seen below. Although he was single author of most of his publications in source journals, he coauthored a number of them with many of those leading physicists (Amaldi, Bethe, Bohr, Frisch, Korff, Landau, Nijboer, Peierls, Teller, Van Hove, Volkoff, and possibly others in wartime classified reports).

## 2. Publications in Source Journals: the G*eneral Search* mode of the WoS

The present version of the *Web of Science* has two search modes: *General Search* mode and *Cited Reference Search* mode. The *General Search* mode archives and reveals publications in source journals starting in the year 1900 (access can be confined to more recent dates depending on financial arrangements). Any of the authors can be used for the search and also combinations thereof. The results are limited to citations (including self-citations) to the papers being queried, appeared in source journals only (not books or popular publications).

A printout of the results of the *General Search* for "Placzek G" (without any other restrictions) is shown in Table 1. The total number of records is rather small. It is even reduced from 22 to 20 when one deletes the two homonyms Gary Placzek (record 14) and Gregory Placzek (record 18). However, the total number of records belonging to George Placzek is increased from 20 to 29 if one adds 9 articles in source journals which seem to have been omitted from the Science Citation Index.

Six of these missing articles appeared in the *Zeitschrift für Physik*, considered a source journal since its inception in 1920 till 1998 when it adopted the name of *European Physics Journal*. However, for reasons unbeknownst to us the issues covering the years 1927 to 1947, i.e. those in which Placzek published in this journal, are not accessible to the WoS. Among the six excluded papers by Placzek is a highly cited one (256 citations) on the rotational structure of Raman bands, coauthored with Edward Teller.[6] Another highly cited article, by L. Landau and G. Placzek (217 citations), dealing with the Landau-Placzek Ratio between Brillouin and Raman scattering, is also missing in Table 1. The reason is that it appeared in German in the *Physikalische Zeitschrift der Sowjetunion* [7], at a time (1934) when Soviet journals are not considered source journals by the ISI.

Record 1 of 22
Author(s): PLACZEK, G
Title: THE SCATTERING OF NEUTRONS BY SYSTEMS OF HEAVY NUCLEI
Source: PHYSICAL REVIEW, 86 (3): 377-388 1952
Times Cited: 453

Record 2 of 22
Author(s): Bethe, HA; Placzek, G
Title: Resonance effects in nuclear processes
Source: PHYSICAL REVIEW, 51 (6): 450-484 MAR 1937
Times Cited: 175

Record 3 of 22
Author(s): PLACZEK, G; VANHOVE, L
Title: CRYSTAL DYNAMICS AND INELASTIC SCATTERING OF NEUTRONS
Source: PHYSICAL REVIEW, 93 (6): 1207-1214 1954
Times Cited: 137

Record 4 of 22
Author(s): PLACZEK, G
Title: ON THE THEORY OF THE SLOWING DOWN OF NEUTRONS IN HEAVY SUBSTANCES
Source: PHYSICAL REVIEW, 69 (9-10): 423-438 1946
Times Cited: 113

Record 5 of 22
Author(s): Bethe, HA; Korff, SA; Placzek, G
Title: On the interpretation of neutron measurements in cosmic radiation
Source: PHYSICAL REVIEW, 57 (7): 573-587 APR 1940
Times Cited: 83

Record 6 of 22
Author(s): PLACZEK, G; SEIDEL, W
Title: MILNE PROBLEM IN TRANSPORT THEORY

Source: PHYSICAL REVIEW, 72 (7): 550-555 1947
Times Cited: 71

Record 7 of 22
Author(s): PLACZEK, G; NIJBOER, BRA; VANHOVE, L
Title: EFFECT OF SHORT WAVELENGTH INTERFERENCE ON NEUTRON SCATTERING BY DENSE SYSTEMS OF HEAVY NUCLEI
Source: PHYSICAL REVIEW, 82 (3): 392-403 1951
Times Cited: 47

Record 8 of 22
Author(s): PLACZEK, G
Title: THE ANGULAR DISTRIBUTION OF NEUTRONS EMERGING FROM A PLANE SURFACE
Source: PHYSICAL REVIEW, 72 (7): 556-558 1947
Times Cited: 46

Record 9 of 22
Author(s): PLACZEK, G
Title: INCOHERENT NEUTRON SCATTERING BY POLYCRYSTALS
Source: PHYSICAL REVIEW, 93 (4): 895-896 1954
Times Cited: 37

Record 10 of 22
Author(s): Bohr, N; Peierls, R; Placzek, G
Title: Nuclear reactions in the continuous energy region
Source: NATURE, 144: 200-201 JUL-DEC 1939
Times Cited: 25

Record 11 of 22
Author(s): PLACZEK, G
Title: INCOHERENT NEUTRON SCATTERING BY POLYCRYSTALS
Source: PHYSICAL REVIEW, 105 (4): 1240-1241 1957
Times Cited: 19

Record 12 of 22
Author(s): PLACZEK, G; VANHOVE, L
Title: INTERFERENCE EFFECTS IN THE TOTAL NEUTRON SCATTERING CROSS-SECTION OF CRYSTALS
Source: NUOVO CIMENTO, 1 (1): 233-256 1955
Times Cited: 18

Record 13 of 22
Author(s): Placzek, G
Title: Concerning light dissapation near the critical point.
Source: PHYSIKALISCHE ZEITSCHRIFT, 31: 1052-1056 1930
Times Cited: 12

Record 14 of 22
Author(s): Spicer, KR; Costa, JE; Placzek, G
Title: Measuring flood discharge in unstable stream channels using ground-penetrating radar
Source: GEOLOGY, 25 (5): 423-426 MAY 1997
Times Cited: 7

Record 15 of 22
Author(s): PLACZEK, G; VOLKOFF, G
Title: A THEOREM ON NEUTRON MULTIPLICATION
Source: CANADIAN JOURNAL OF RESEARCH SECTION A-PHYSICAL SCIENCES, 25 (4): 276-292 1947
Times Cited: 4

Record 16 of 22
Author(s): PLACZEK, G
Title: THEORY OF SLOW NEUTRON SCATTERING
Source: PHYSICAL REVIEW, 75 (8): 1295-1295 1949
Times Cited: 3

Record 17 of 22
Author(s): PLACZEK, G
Title: CORRECTION
Source: NUOVO CIMENTO, 1 (5): 967-967 1955
Times Cited: 1

Record 18 of 22
Author(s): Placzek, G; Wiser, R; Roberts, KP
Title: Microscopic characterization of CDSE/ZNS nanocrystals.
Source: ABSTRACTS OF PAPERS OF THE AMERICAN CHEMICAL SOCIETY, 229: U391-U392 340-CHED Part 1 MAR 13 2005
Times Cited: 0

Record 19 of 22
Author(s): PLACZEK, G
Title: CORRECTION
Source: PHYSICAL REVIEW, 94 (6): 1801-1801 1954
Times Cited: 0

Record 20 of 22
Author(s): PLACZEK, G
Title: SCATTERING OF X-RAYS BY ATOMS
Source: PHYSICAL REVIEW, 86 (4): 588-588 1952
Times Cited: 0

Record 21 of 22
Author(s): Bethe, HA; Korff, SA; Placzek, G
Title: On the interpretation of neutron measurements in cosmic radiation.
Source: JOURNAL OF THE FRANKLIN INSTITUTE, 230: 776-777 JUL-DEC 1940

Times Cited: 0

Record 22 of 22
Author(s): Placzek, G
Title: Evidence for the spin of the photon from light scattering
Source: NATURE, 128: 410-410 JUL-DEC 1931
Times Cited: 0

**Table 1.** Publications in source journals and their citations listed in the WoS (*General Search* mode) under "Placzek G". Note that records 14 and 18 belong to homonyms.

Two additional articles by Placzek are missing pathologically from Table 1. One of them, on the Raman Effect of gaseous ammonia, appeared in the *Zeitschrift für Physik* [8] (10 citations) coauthored with E. Amaldi. The other, on the capture of slow neutrons, appeared in *Nature* [9], coauthored with O.R. Frisch (7 citations). It does not appear in the WoS under *General Search*. The whole volume 137 of *Nature* is actually missing. Such errors unfortunately are not uncommon for publications which predate 1945 and thus have been added to the WoS only recently (in 2005). Such publications can be retrieved in the *Cited Refernce Search* mode (see next section). The total number of citations received by these 29 papers is 1734. This includes self citations but it does not include incorrect citations which, as we well see next, can be often revealed by using the *Cited Reference Search* mode of the WoS.

## 3. References to publications which did not appear in source journals or were incorrectly cited: The *Cited Reference Search* mode of the WoS

The *Cited Reference Search* mode enables access to all references appeared in source journals (whether the references are to articles in source journals cited either correctly or incorrectly), or references to articles published in non-source journals or in books or any other published material (sometimes even unpublished, e.g., theses, internal reports, or even private communications). Probably the most useful feature of this mode, especially concerning Placzek's work, is the possibility of finding a measure of the impact of a book or a book article, as reflected in the citations in source journals. Conference proceedings, even if not published in source journals, can also be accessed using the *Cited Reference Search* mode.

As an example we deal with the problem of the missing Ref. 9. Because the articles in this missing volume of *Nature* are not source items, only the first author (Frisch O R ) must be queried under the *Cited Reference Search*. We then find 7 citations for Ref. 9. Care must be taken searching either O R or O* and not OR (without blank). If the latter is done the system takes OR as a Boolean operator and gives an error message. The same procedure can be used to find the citations of Ref. 7. One must take into consideration here that the name of the first author deviates from the standard form "Landau L D" and the citations (217 citations) must be retrieved either under "Landau L" or "Landau L*".

Placzek published in 1934 in the *Handbuch der Radiologie* a 270 pages article under the title (in German): "Rayleigh scattering and the Raman Effect". It is probably the first comprehensive article on the theory of light scattering published after the discovery of the Raman Effect in 1928. In spite of having been written in German at a time when Germany's star was beginning to wane, a *Cited Reference Search* reveals a total of about 986 citations for this article (see Table 2).

Column 1: Consecutive numbering
Column 2: Number of citations at the date of search
from Column 3: Short form of the cited publications

```
    1    453    PLACZEK G       1952   V86    P377      PHYS REV
    2    360    PLACZEK G       1934   V6     P205      HDB RADIOLOGIE
    3    280    PLACZEK G       1934   V6     P209      HANDB RADIOL
    4    254    PLACZEK G       1933   V81    P209      Z PHYS
    5    174    ...Placzek G    1937   V51    P450      PHYS REV
    6    137    PLACZEK G       1954   V93    P1207     PHYS REV
    7    113    PLACZEK G       1946   V69    P423      PHYS REV
    8     83    ...Placzek G    1940   V57    P573      PHYS REV
    9     83    PLACZEK G       1934   V2     P209      HDB RADIOLOGIE
   10     71    PLACZEK G       1947   V72    P550      PHYS REV
   11     47    PLACZEK G       1931   V70    P84       Z PHYS
   12     47    PLACZEK G       1951   V82    P392      PHYS REV
   13     46    PLACZEK G       1947   V72    P556      PHYS REV
   14     37    PLACZEK G       1954   V93    P895      PHYS REV
   15     36    PLACZEK G       1940   V57    PA1075    PHYS REV
   16     27    PLACZEK G       1934   V6               HANDBUCH RADIOLOGI
   17     25    ...Placzek G    1939   V144   P200      NATURE
   18     24    PLACZEK G       1934   V6     P71       HDB RADIOLOGIE
   19     20    PLACZEK G       1934   V6     P2        HDB RADIOLOGIE
   20     19    PLACZEK G       1957   V105   P1240     PHYS REV
   21     18    PLACZEK G       1955   V1     P233      NUOVO CIMENTO
   22     15    PLACZEK G       1946   V37    P57       NRC1547 NAT RES COU
   23     14    PLACZEK G       1930   V31    P1052     PHYS Z
   24     14    PLACZEK G       1931   V1     P71       LEIPZIGER VORTRAGE
   25     12    PLACZEK G       1929   V58    P585      Z PHYS
   26     12    PLACZEK G       1935   V4     P209      RAYLEIGH SCATTERING
   27     11    PLACZEK G       1934   V2     P205      HDB RADIOLOGIE 6
   28     11    PLACZEK G       1995          P133      A25 MANH PROJ REP
   29     10    PLACZEK G       1950          P581      2 BERK S MATH STAT
   30      9    PLACZEK G       1946   V25    P209      FUNCTIONS ENX
   31      9    PLACZEK G       1959   V6     P526      RAYLEIGH RAMAN SCAT
   32      9    PLACZEK G       1962   V6     P109      RAYLEIGH RAMAN SCAT
   33      8    PLACZEK G       1947   V25    P276      CANADIAN J RES A
   34      8    PLACZEK G       1951          P581      2 P BERK S MATH STA
   35      8    PLACZEK G       1959   V4     P209      RAYLEIGH SCATTERING
```

| | | | | | | |
|---|---|---|---|---|---|---|
| 36 | 8 | PLACZEK G | 1962 | | P139 | UCRLTRANS526L US AT |
| 37 | 7 | ...Placzek G | 1997 | V25 | P423 | GEOLOGY |
| 38 | 6 | PLACZEK G | 1928 | V49 | P601 | Z PHYS |
| 39 | 6 | PLACZEK G | 1932 | V2 | P91 | STRUCTURE MOLECULES |
| 40 | 6 | PLACZEK G | 1934 | V4 | P205 | HANDBUCH RADIOLOGIE |
| 41 | 6 | PLACZEK G | 1934 | V6 | P203 | MARX HDB RADIOLOGIE |
| 42 | 6 | PLACZEK G | 1946 | | P49 | MT1 NAT RES COUNC C |
| 43 | 6 | PLACZEK G | 1962 | V526 | P175 | UCRL TRANS |
| 44 | 5 | PLACZEK G | 1929 | V55 | P81 | Z PHYS |
| 45 | 5 | PLACZEK G | 1931 | V1 | P293 | LEIPZIGER VORTRAGE |
| 46 | 5 | PLACZEK G | 1931 | V67 | P582 | Z PHYS |
| 47 | 5 | PLACZEK G | 1934 | V6 | P206 | MARX HDB RADIOLOGIE |
| 48 | 5 | PLACZEK G | 1941 | | P133 | A25 MANH PROJ REP |
| 49 | 5 | PLACZEK G | 1947 | | P6 | MT16 |
| 50 | 4 | PLACZEK G | 1934 | V6 | P365 | HDB RADIOLOGIE |
| 51 | 4 | PLACZEK G | 1934 | V6 | PCH3 | HDB RADIOLOGIE 2 |
| 52 | 4 | PLACZEK G | 1934 | V6 | PR4 | HANDBUCH RADIOLOGIE |
| 53 | 4 | PLACZEK G | 1936 | V6 | P293 | IN PRESS PHYS REV |
| 54 | 4 | PLACZEK G | 1943 | | P6 | MT4 |
| 55 | 3 | PLACZEK G | 1931 | V1 | P100 | LEIPZIGER VORTRAGE |
| 56 | 3 | PLACZEK G | 1932 | V2 | P65 | STRUCTURE MOLECULES |
| 57 | 3 | PLACZEK G | 1934 | V4 | P206 | MARX HUNDBUCH RADIO |
| 58 | 3 | PLACZEK G | 1934 | V4 | P209 | HDB RADIOLOGIE |
| 59 | 3 | PLACZEK G | 1934 | V4 | P366 | HDB RADIOLOGIE |
| 60 | 3 | PLACZEK G | 1934 | V4 | P371 | HANDBUCH RADIOLOGI |
| 61 | 3 | PLACZEK G | 1934 | V6 | P109 | RAYLEIGH RAMAN SCAT |
| 62 | 3 | PLACZEK G | 1934 | V6 | P366 | HDB RADIOLOGIE |
| 63 | 3 | PLACZEK G | 1934 | V6 | P423 | HANABUCH RADIOLOGIE |
| 64 | 3 | PLACZEK G | 1934 | V6 | P64 | RAYLEIGH RAMAN SCAT |
| 65 | 3 | PLACZEK G | 1934 | V6 | PCH12 | HANDBUCH RADIOLOGY |
| 66 | 3 | PLACZEK G | 1934 | V7 | P203 | MARX HDB RADIOLOGIE |
| 67 | 3 | PLACZEK G | 1946 | | P6 | MT1 NAT RES COUNC D |
| 68 | 3 | PLACZEK G | 1949 | V75 | P1295 | PHYS REV |
| 69 | 3 | PLACZEK G | 1951 | V37 | P57 | NRC1547 |
| 70 | 3 | PLACZEK G | 1954 | V37 | P57 | NBS APPL MATH SER |
| 71 | 3 | PLACZEK G | 1954 | V93 | P897 | PHYS REV |
| 72 | 3 | PLACZEK G | 1959 | | P133 | 526 UCRL |
| 73 | 3 | PLACZEK G | 1962 | V526 | P138 | UCRLT526 L REP |
| 74 | 3 | PLACZEK G | 1980 | | P49 | MOL VIBRATIONS |
| 75 | 2 | PLACZEK G | 1929 | V38 | P585 | Z PHYS |
| 76 | 2 | PLACZEK G | 1933 | V81 | P839 | Z PHYSIK |
| 77 | 2 | PLACZEK G | 1933 | V83 | P209 | Z PHYS |
| 78 | 2 | PLACZEK G | 1934 | | | HANDB RADIOLOGIE |
| 79 | 2 | PLACZEK G | 1934 | | P139 | UCRLTRANS526L DEP C |
| 80 | 2 | PLACZEK G | 1934 | V12 | P209 | HDB RADIOLOGIE |
| 81 | 2 | PLACZEK G | 1934 | V2 | P316 | HDB RADIOLOGIE |
| 82 | 2 | PLACZEK G | 1934 | V2 | P328 | HDB RADIOLGIE |
| 83 | 2 | PLACZEK G | 1934 | V2 | P343 | HDB RADIOLOGIE 4 |
| 84 | 2 | PLACZEK G | 1934 | V6 | P208 | HDB RADIOLOGIE |
| 85 | 2 | PLACZEK G | 1934 | V6 | P224 | HDB RADIOLOGIE |
| 86 | 2 | PLACZEK G | 1934 | V6 | P276 | HDB RADIOLOGIE |
| 87 | 2 | PLACZEK G | 1934 | V6 | P283 | HDB RADIOLOGIE |
| 88 | 2 | PLACZEK G | 1934 | V6 | P293 | HDB RADIOLOGIE |
| 89 | 2 | PLACZEK G | 1934 | V6 | P321 | HDB RADIOLOGIE 2 |
| 90 | 2 | PLACZEK G | 1934 | V6 | P323 | HDB RADIOLOGIE 2 |
| 91 | 2 | PLACZEK G | 1934 | V6 | P355 | MARX HDB RADIOLOGI |
| 92 | 2 | PLACZEK G | 1934 | V6 | P371 | HANDBUCH RADIOLOGI |
| 93 | 2 | PLACZEK G | 1935 | V2 | P209 | HDB RADIOLOGIE |
| 94  | 2 | PLACZEK G | 1935 | V6   | P209  | HDB RADIOLOGIE |
| 95  | 2 | PLACZEK G | 1938 | V83  | P209  | Z PHYS |
| 96  | 2 | PLACZEK G | 1939 | V6   | P209  | HDB RADIOLOGIE |
| 97  | 2 | PLACZEK G | 1940 | V57  | P1072 | PHYS REV |
| 98  | 2 | PLACZEK G | 1942 |      | P133  | A25 MANH PROJ REP |
| 99  | 2 | PLACZEK G | 1947 |      | P49   | MT1 AE PROJ NAT RES |
| 100 | 2 | PLACZEK G | 1947 | V72  | P50   | PHYS REV |
| 101 | 2 | PLACZEK G | 1949 | V2   | PCH7  | SCI ENG NUCLEAR POW |
| 102 | 2 | PLACZEK G | 1950 |      | P281  | UNPUB P BERKELEY S |
| 103 | 2 | PLACZEK G | 1953 | V71  | PS100 | LOS ALAMOS SCI LAB |
| 104 | 2 | PLACZEK G | 1956 | V37  | P57   | NOMT1 NAT RES COUNC |
| 105 | 2 | PLACZEK G | 1959 | V526 | P138  | UCRLTRANS526 U CAL |
| 106 | 2 | PLACZEK G | 1959 | V6   | P139  | RAYLEIGH RAMAN SCAT |
| 107 | 2 | PLACZEK G | 1962 |      |       | UCRLTRANS526L |
| 108 | 2 | PLACZEK G | 1962 |      | P175  | UCRLTRANS526L US AT |
| 109 | 2 | PLACZEK G | 1962 | V526 | P562  | UCRL T |
| 110 | 2 | PLACZEK G | 1980 |      |       | MOL VIBRATIONS |
| 111 | 2 | PLACZEK G | 1995 |      |       | MOL VIBRATIONS |
| 112 | 1 | PLACZEK G |      |      |       | |
| 113 | 1 | PLACZEK G |      |      |       | |
| 114 | 1 | PLACZEK G |      |      |       | |
| 115 | 1 | PLACZEK G | 1925 | V6   | P205  | HDB RADIOLOGIE 2 |
| 116 | 1 | PLACZEK G | 1929 | V81  | P81   | Z PHYS |
| 117 | 1 | PLACZEK G | 1930 | V31  | P1051 | PHYS Z |
| 118 | 1 | PLACZEK G | 1930 | V33  | P832  | P AMSTERDAM |
| 119 | 1 | PLACZEK G | 1930 | V38  | P832  | P AMST |
| 120 | 1 | PLACZEK G | 1930 | V67  | P582  | Z PHYS |
| 121 | 1 | PLACZEK G | 1931 | V1   | P105  | LEIPZIGER VORTRAGE |
| 122 | 1 | PLACZEK G | 1931 | V1   | P75   | LEIPZIGER VORTRAGE |
| 123 | 1 | PLACZEK G | 1931 | V1   | P81   | LEIPZIGER VORTRAGE |
| 124 | 1 | PLACZEK G | 1931 | V1   | P94   | LEIPZIGER VORTRAGE |
| 125 | 1 | PLACZEK G | 1931 | V1   | P96   | LEIPZIGER VORTRAGE |
| 126 | 1 | PLACZEK G | 1931 | V1   | PS100 | LEIPZIGER VORTRAGE |
| 127 | 1 | PLACZEK G | 1931 | V70  | P287  | Z PHYS |
| 128 | 1 | PLACZEK G | 1931 | V70  | P83   | Z PHYS |
| 129 | 1 | PLACZEK G | 1931 | V70  | PS4   | Z PHYS |
| 130 | 1 | PLACZEK G | 1931 | V71  | PS100 | LEIPZIGER VORTRAGE |
| 131 | 1 | PLACZEK G | 1931 | V72  | P257  | Z PHYS |
| 132 | 1 | PLACZEK G | 1932 | V4   | P211  | HDB RADIOLOGIE |
| 133 | 1 | PLACZEK G | 1932 | V6   | P2    | HDB RADIOLOGIE |
| 134 | 1 | PLACZEK G | 1932 | V6   | P211  | HDB RADIOLOGIE |
| 135 | 1 | PLACZEK G | 1932 | V6   | P339  | HDB RADIOLOGIE |
| 136 | 1 | PLACZEK G | 1933 | V6   | P2    | HDB RADIOLOGIE |
| 137 | 1 | PLACZEK G | 1933 | V6   | P205  | MARX HANDBUCH RADI |
| 138 | 1 | PLACZEK G | 1933 | V6   | P339  | HDB RADIOLOGIE |
| 139 | 1 | PLACZEK G | 1933 | V81  | P201  | Z PHYS |
| 140 | 1 | PLACZEK G | 1933 | V81  | P208  | Z PHYS |
| 141 | 1 | PLACZEK G | 1934 |      | P205  | ACADEISCHE VERLAG |
| 142 | 1 | PLACZEK G | 1934 | V1   | P160  | HDB RADIOLOGIE |
| 143 | 1 | PLACZEK G | 1934 | V1   | P244  | HDB RADIOLOGIE 2 |
| 144 | 1 | PLACZEK G | 1934 | V2   | P244  | HDB RADIOLOGIE 2 |
| 145 | 1 | PLACZEK G | 1934 | V2   | P308  | HDB RADIOLOGIE |
| 146 | 1 | PLACZEK G | 1934 | V2   | P365  | HDB RADIOLOGIE |
| 147 | 1 | PLACZEK G | 1934 | V2   | PCH12 | HANDBUCH RADIOLOGIE |
| 148 | 1 | PLACZEK G | 1934 | V2   | PCH21 | HDB RADIOLOGIE |
| 149 | 1 | PLACZEK G | 1934 | V25  | P209  | DANS HDB RADIOLOGIE |
| 150 | 1 | PLACZEK G | 1934 | V25  | P423  | HANABUCH RADIOLOGIE |
| 151 | 1 | PLACZEK G | 1934 | V4   | P238  | HANDBUCH RADIOLOGIE |

```
210     1    PLACZEK G     1951            P69      2 P S MATH STAT PRO
211     1    PLACZEK G     1952    V31     P377     PHYS REV
212     1    PLACZEK G     1952    V86     P624     PHYS REV
213     1    PLACZEK G     1952    V86     P630     PHYS REV
214     1    PLACZEK G     1952    V88     P377     PHYS REV
215     1    PLACZEK G     1953    V1      P293     INTRO THEORY NEUTRO
216     1    PLACZEK G     1953    V93     P1027    PHYS REV
217     1    PLACZEK G     1954    V144    P57      NBS APPL MATH    3
218     1    PLACZEK G     1954    V37     P281     US NBS APPL MATH SE
219     1    PLACZEK G     1954    V82     P392     PHYS REV
220     1    PLACZEK G     1954    V93     P1027    PHYS REV
221     1    PLACZEK G     1954    V93     P1212    PHYS REV
222     1    PLACZEK G     1954    V93     P352     PHYS REV
223     1    PLACZEK G     1954    V93     P595     PHYS REV
224     1    PLACZEK G     1954    V93     PL895    PHYS REV
225     1    PLACZEK G     1957    V9      P567     Z PHYSIK
226     1    PLACZEK G     1959    V526    P175     UCRL526L
227     1    PLACZEK G     1959    V526    PCH8     UCRL526L T
228     1    PLACZEK G     1959    V6      P186     RAYLEIGH RAMAN SCAT
229     1    PLACZEK G     1959    V6      P206     RAYLEIGH RAMAN SCAT
230     1    PLACZEK G     1959    V6      P209     REYLEIGH SCATTERING
231     1    PLACZEK G     1959    V6      P211     HDB RADIOLOGIE
232     1    PLACZEK G     1960    V4      P209     RAYLEIGH SCATTERING
233     1    PLACZEK G     1962            P133     526 U CAL RAD LAB T
234     1    PLACZEK G     1962            P562     TRANSL NEWS
235     1    PLACZEK G     1962            P97      UCRLTRANS526L US AT
236     1    PLACZEK G     1962    V2      P65      T526L UCRL US AT EN
237     1    PLACZEK G     1962    V256    P209     RAYLEIGH RAMAN STRU
238     1    PLACZEK G     1962    V37     P562     USAAEC UCRL T
239     1    PLACZEK G     1962    V524    P562     UCRL T
240     1    PLACZEK G     1962    V526    P205     UCRL526L TRANSL
241     1    PLACZEK G     1962    V526    PCH8     UCRL526L REP
242     1    PLACZEK G     1962    V562    P2       USAAEC UCRL T
243     1    PLACZEK G     1962    V6      P209     HDB RADIOLOGIE 2
244     1    PLACZEK G     1962    V6      P526     RAYLEIGH RAMAN SCAT
245     1    PLACZEK G     1964    V69     P423     PHYS REV
246     1    PLACZEK G     1972    V57     P2264    Z PHYS
247     1    PLACZEK G     1982            P281     US AEC UCRLTRANS526
248     1    PLACZEK G     1995            P49      MOLEC VIBRATIONS
```

Entries 112-114: No details about the corresponding references are given. This must be either an oversight or corresponds to private communications.

**Table 2.** Placzek's publications as appear in the WoS within the *Cited Reference Search* mode. Some of the most obvious cases of reference journal variations have been bundled together. The observant reader will still discover many reference errors.

We now illustrate on the basis of Table 2 the evaluation of citations to books and book chapters by articles in source journals. The second most cited entry in this table is listed to have been cited 360 times, having appeared in *Handbuch der Radiologie* in 1934 (Vol. 6, page 205). The third entry (280 citations) corresponds to the same publication but citing page 209 (and thus appears as if it were a wholly different publication). The 9$^{th}$ entry (83 citations) corresponds obviously to the same publication but the volume is erroneously given as the 2$^{nd}$, instead of the 6$^{th}$. Hence, although it is the same, it is listed as a fully different publication. A total citation count for this book chapter must be performed by hand and include also the entries 16 (no page given, 27 citations), 18 (page 71, 24 citations), etc. By now it should have become clear that some authors list in the reference the first page of the chapter while others give simply the page that they are using. For a 270 page article the possibility of errors is rather large. The cumulative citations to the *Handbuch der Radiologie* article are obtained by hand from Table 2 to be 986. This is to be regarded as a lower limit corresponding to a single citation per citing article. Since a large article, like the one at hand, may be cited several times in a citing article, the total number of citations may be somewhat higher.

The outlined procedure can also be applied to the book on neutron diffusion, coauthored by Placzek.[10] Although the book is known to be highly cited, it appears only once on Table 2 (see entry 215 in the Table), where the first author was incorrectly given as Placzek. The reason for the missing citations seems to be that Placzek is not the correct first author of the book and is thus often not cited in connection with it. Searching for the citations to the first author K.M. Case, in the *Cited Reference Search* mode one finds many to this book in several variations. The sum of all of them leads to a total of 630. The remaining items in Table 2 correspond to erroneous citations, internal reports, especially declassified ones related to the Manhattan project, lecture notes and, last but not least, citations of work appeared in non-source journals. The *Cited Reference Search* mode allows a counting of all references in source items, yielding for Table 2 a total of 2876. The 630 citations to Ref. 10 should also be added. This gives us a lower limit for the total number of citations of 2876 + 630 = 3506.

In order to clarify the rather complicated pattern of the results in Table 2 we show in Figure 1 a bar diagram representing the number of citations received by Placzek's articles, books, reports, etc. versus the year of publication. His first article, published in the *Zeitschrift für Physik* in 1928 and cited 6 times, is based on his doctoral thesis performed at the University of Vienna (title: Determination of density and shape of submicroscopic test bodies). His second article, published in 1929, shortly after the discovery of the Raman Effect, dealt with the theory of the Raman Effect. The years 1933-1934 represent his "annus mirabilis", with nearly 1500 citations. During these years he published the article in the *Handbuch der Radiologie*, the work on the rotational spectra of molecules (with E. Teller) and the theory of the Landau-Placzek Ratio. In 1952 his most cited article (454 citations) appeared in the *Physical Review*, a journal that after emigration took the place of the *Zeitschrift für Physik* for communicating his work. The title of the paper, of which he was single author, was "The scattering of neutrons by systems of heavy nuclei". During 1952 he must also have been writing the book in Ref. 10.

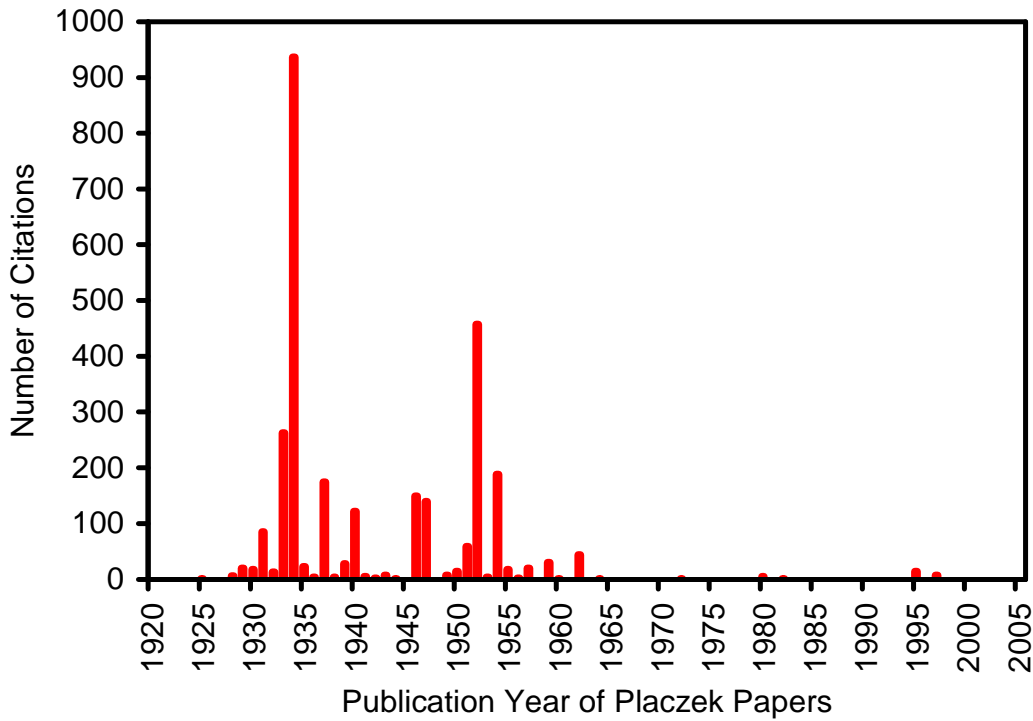

**Figure 1.** Citation bar diagram with the number of citations vs. publication year of the Placzek papers as listed in Table 2. Note that his contribution to the book in Ref. 10 (630 citations) is not included for the reasons given in the text. For the meaning of "posthumous" papers see text.

According to Figure 1 there seem to be a few posthumous publications after his untimely death in 1955. The most recent one (Geology, 1997) is not his work but that of a namesake. However, that listed as published in 1995 corresponds to a report of the Manhattan project, cited 5 times (see Table 2) but hard to obtain for ordinary mortals. The 1997 paper (7 citations) is actually a publication by Gary Placzek, who worked at the US geologic survey. We have succeeded in contacting the other namesake who appears as the 18[th] record in Table 1 (Gregory Placzek). He is an undergraduate at the University of Tulsa, majoring in chemistry and is not related to George (only a few members of the Brno Placzek clan managed to survive the holocaust). The reason why he does not appear in Table 2 or in Figure 1 is that his paper, an abstract of a talk delivered this year, has no citations (yet). The user of the WoS should keep in mind that uncited papers appear in the *General*

*Search* mode but not under the *Cited Reference Search* mode (after all, they have not been cited yet).

The really posthumous publication by George Placzek was submitted in 1956 (Incoherent neutron scattering by polycrystals), indeed after Placzek's death.[11] It is based on sketchy, unpublished notes which, because of their interest, were edited and refereed by L. Van Hove, a rather unusual procedure justified only by the importance of the work (number 20 in Table 2, 19 citations). Papers No. 31 (1959) and 32 (1962) in Table 2 and Figure 1 seem to refer to some English translation of paper No. 2, like many of his "pseudo-posthumous" papers unavailable to us.

We include next a figure with the time evolution of the yearly citations of Placzek's most cited papers (1, 2-3, and 4 of Table 2). The citations to the *Handbuch* article (1993) increased drastically from 1960 till 1977, reaching the rather high value of 40 citations per year in 1977. They have been falling smoothly since then, to reach the still important value of 13 in 2005 (many authors would be happy to be cited as frequently 70 years after publication). The decrease may be due to the appearance of a large number of books on light scattering.[12] Let us not forget, however, that the book was written in German and no English translation, if extant, is easily available. The citations to the Placzek and Teller article [6] (No. 4 of Table 2) follow a similar time evolution, with the vertical axis properly scaled. It was also written in German which, by 1975, had lost its appeal as a scientific language. The yearly citations of Placzek's most cited paper, however, after rising to a maximum around 1975, have remained nearly constant at around 10 per year.

15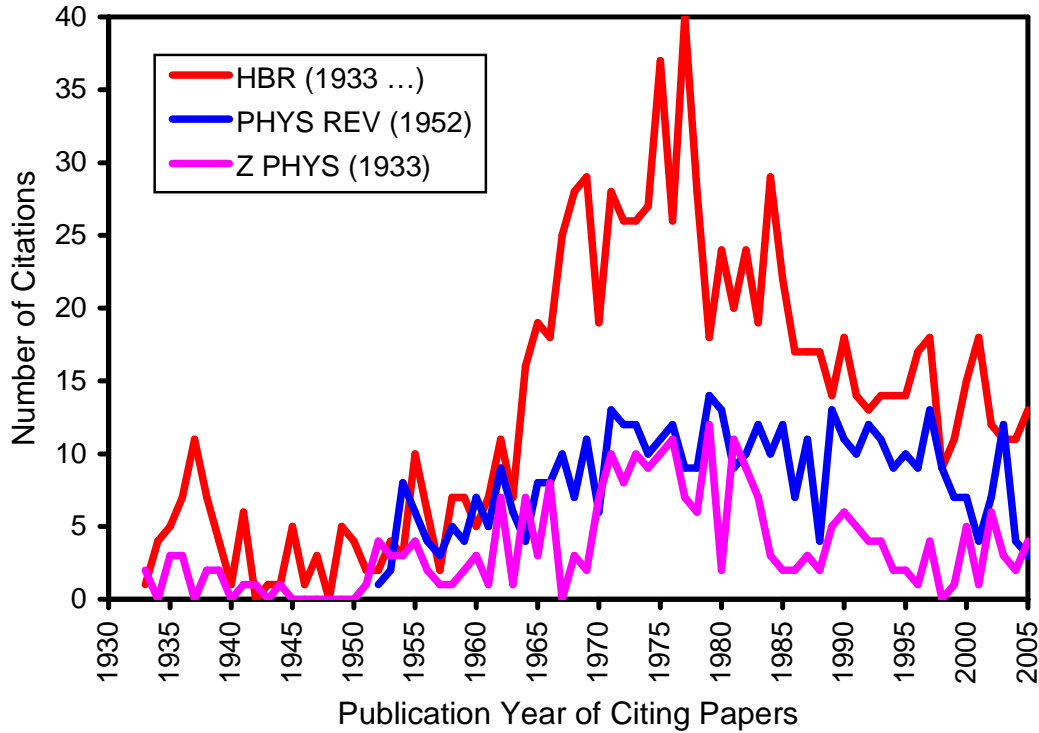

**Figure 2.** Citation history of Placzek's three most cited papers. HBR represents the *Handbuch der Radiologie*.

What else does Figure 2 tell us? The number of citations received by the three publications remains nearly constant at a low level till about 1955-1960. They take off sometime within this period to reach their maxima around 1975. Because of the long period of low citation activity, followed by a big upsurge, they deserve the name of *sleeping beauties*.[13] One may speculate about the causes for this relatively sharp growth in the late fifties. We mention two possibilities: the *Sputnik Effect* and the advent of Laser-Raman spectroscopy and also inelastic neutron scattering. The launching of *Sputnik* in 1957 produced a shock in the United States which was to be counteracted by massive support and development in the physical science and engineering. Raman spectroscopy had remained a rather academic specialty since its discovery in 1928, because of the weak light sources available. The invention and industrial production of gas lasers resulted in easy accessibility of Raman spectroscopy during the mid sixties. The colossal development that followed brought to the fore Placzek's *Handbuch* article at a time when hardly any in-depth publications of the underlying theory were available.

The time dependence pattern shown in Figure 2 for the three most cited of Placzek's works is also found for the sum of all citations (cumulative impact). This is illustrated in Figure 3 which indicates that his work is still cited at the rate of about 35 citing papers per year, a rather remarkable number of citations for publications that, on the average, are 75 years old.[14]

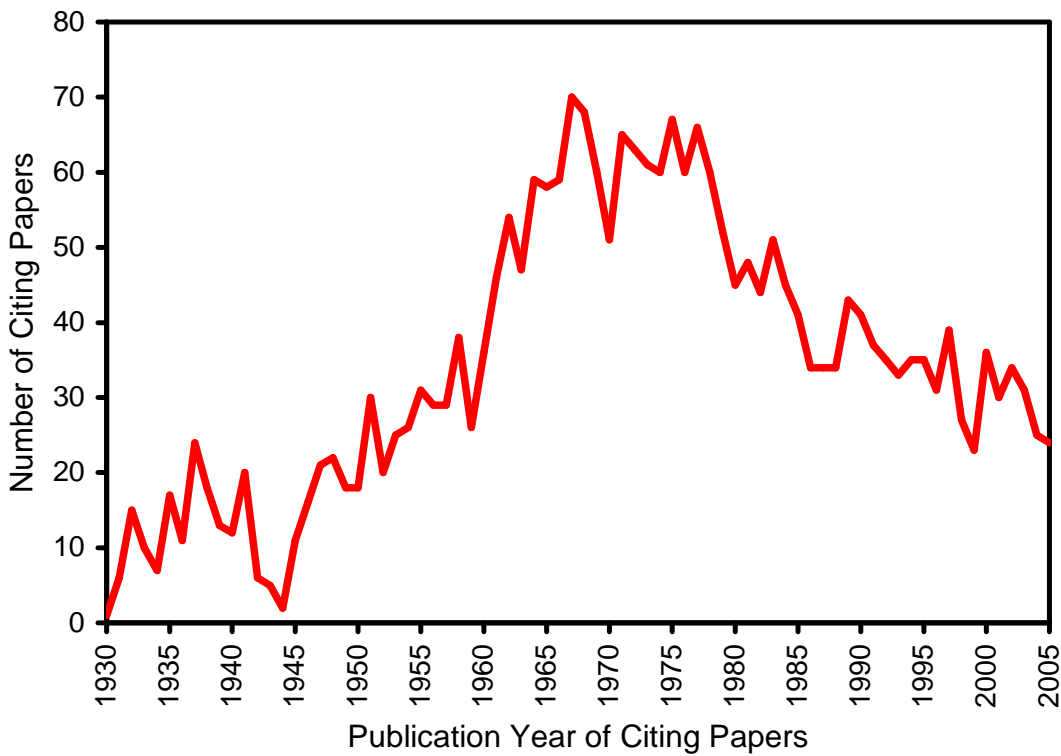

**Figure 3.** Time dependence of the citations of the ensemble of all Placzek publications. The number of **citing** papers versus their year of publication is given. The 630 citations of the book by Case and Placzek are not included here.

**4. Informal citations by name only**

Some contributions to science become household words (e.g. Maxwell´s Equations, Bose-Einstein Statistics, the Ginzburg-Landau Equations, the Landau-Placzek Ratio) and full reference to the original work, in terms of a formal citation, is then often not given. Instead, the name of the author or the concept bearing his name is given, often in the title, the abstract or the keywords. It then becomes difficult for those of us who want to go to the original sources to figure out where to find them. This widespread practice may also pose problems for settling priority issues. Fortunately, the WoS, in its *General Search* mode, querying for the name of the author as a topic offers the possibility of finding out how many times the author is mentioned as an "informal reference" provided his name appears in the title of a paper, the abstract, or the keywords.[15] We have found that the name "Placzek" has been thus mentioned informally in 118 publications.

It is also possible with the Boolean operators available in the WoS to find out in what connection a given author is cited informally. It suffices to look at the title, the abstract and/or the keywords of one of the informally citing papers to find out one associated concept and then query under "topic" the author's name followed by the Boolean operator AND and the desired concept. The results found for Placzek are:

| | |
|---|---|
| Landau-Placzek Ratio | 61 times |
| Placzek's approximation | 28 times |
| Placzek's polarizability | 17 times |
| Placzek's lemma | 9 times |
| Placzek's theory | 2 times |

Although neither Placzek's approximation nor Placzek's theory are very tale telling terms, the Landau-Placzek Ratio, Placzek's polarizability and Placzek's lemma have become household words.

We display in Table 3 the number of informal citations for Placzek and his prominent coworkers. In all cases except for Frisch and Case it can be easily found from the WoS, *General Search* mode (under *Topic*). No such results are given for either Frisch, Korff or Case. For Frisch and Korff, the large number of namesakes in the biomedical field would impose a rather tedious separation by hand. For Case, the various ordinary meanings of the word do not allow the computer to recognize it as a family name. For comparison we have added to Table 3 C.V. Raman, Raman spectroscopy being one of the fields in which Placzek excelled, and two living Nobel laureates, Ginzburg and Hänsch.

| Author | Informal citations | Reference citations | Hirsch number h | Andersen index IA |
|---|---|---|---|---|
| E. Amaldi (1931-1989) | 46 | 1331 | 19 | 0.40 |
| H. Bethe * (1927-2004) | 5616 | 20504 | 58 | 3.46 |
| N. Bohr * (1909-1962) | 2938 | 2716 | 20 | 0.97 |
| K.M. Case (1948-1989) |  | 3034 | 27 | 1.80 |
| S.A. Korff (1927-1982) |  | 911 | 13 | 0.30 |
| L.D. Landau * (1926-1968) | 19155 | 10919 | 31 | 6.19 |
| B.R.A. Nijboer (1937-1988) | 19 | 1488 | 16 | 0.57 |
| R. Peierls (1929-1995) | 4442 | 5556 | 28 | 1.04 |
| G. Placzek (1930-1955) | 118 | 1252 | 12 | 2.00 |
| E. Teller (1931-2003) | 9661 | 27742 | 40 | 5.35 |
| L. Van Hove (1949-1990) | 297 | 9913 | 43 | 5.90 |
| G.M. Volkoff (1939-1963) | 67 | 1128 | 11 | 1.96 |
| C.V. Raman * (1907-1970) | 95669 | 1152 | 17 | 0.29 |
| V.L. Ginzburg * (1939-) | 5960 | 21243 | 40 | 4,88 |
| T.W. Hänsch * (1969-) | 300 | 12405 | 62 | 9.57 |

* Nobel Laureates

**Table 3.** Several impact indexes of Placzek, some of his coworkers and, for comparison, C.V. Raman and two living Nobel laureates (Ginzburg and Hänsch). The informal references are discussed in Sect. 4, the h [16] and IA [17] indexes in Sects. 5 and 6, respectively. In parentheses: time periods of scientific activity.

Table 3 exposes Landau and Raman as the two giants of the informal citations, followed by Teller and Bethe. Landau is mostly informally cited in connection with the Ginzburg-Landau Equations (Ginzburg: Nobel Laureate 2003, Landau: NL 1962). Bethe (NL 1967) is mostly cited informally in connection with the Bethe Ansatz. Raman (NL 1930), of course becomes a household word in connection with Raman spectroscopy, Placzek in connection with the Landau-Placzek Ratio.

It is interesting to note that there is no direct correlation between the informal citations and the two impact indexes h and IA to be discussed in the next two sections. There may actually be some kind of anti-correlation: informal citations are often used at the expense of the corresponding formal ones.

## 5. The Hirsch index h

This index was introduced recently by J. Hirsch as a measure of the cumulative impact of a person's scientific work within a given discipline.[16] It can be easily obtained, within a minute or so, provided one has access to the WoS (*General Search* mode) and the number of highly cited namesakes is low. The index h is simply defined as the number of articles in source journals that have had h citations or more. It is not easy to include books and other items listed in the *Cited Reference Search* mode (we do not do it here). The index h increases with the age of the scientist and comparison between different disciplines should be avoided unless a reliable relative calibration is available.[18]

Particularly striking is the comparison of the informal references of Raman (a whopping 95669) with his low h (18, only slightly higher than that of Placzek). The total number of Raman's citations (1570 as obtained in the *Cited Reference Search* mode) is also considerably lower than that found for Placzek (2627). It is easy to conjecture the reason for this apparent discrepancy. The Raman Effect was discovered in 1928 and awarded the Nobel Prize in 1930, which made it almost immediately a household word. The original source item publication appeared in *Nature*[19] but was cited only 170 times: authors preferred to mention simply Raman Effect than give a formal citation. Many of his publications appeared in Indian journals which are even now not in the list of source items. As an example we mention an article in the *Proc. of the Indian Acad. of Sciences* **2**, 406 (1935), cited 383 times (Raman's most cited paper). Later issues of this journal are now listed as source items.

We conclude the discussion of Table 3 by mentioning E. Amaldi, a close friend and coauthor of Placzek. The small number of informal citations corresponds mainly to the "Fermi-Amaldi Correlation Terms". Amaldi's Hirsch number is only slightly higher than that of Placzek inspite of his having been blessed with a much longer life (1908-1989). In principle, the impact of early papers and thereby the overall impact of pioneers like Niels Bohr are highly underestimated: The proliferation of science implies a proliferation of citable papers, resulting in increasing ratios of references per paper (reference count) and citations per paper, respectively. These ratios have approximately doubled within the last half century. Hence, the Hirsch numbers of scientists from different time periods are hardly comparable.

**6. The Andersen Index IA**

The index IA, proposed by O.K. Andersen [17], is obtained with the following expression:

$$IA = \frac{\text{number of citations}}{(\text{years of scientific life})^2}$$

In the above expression the number of citations is usually obtained from the *General Search mode* of the WoS, although a correction for wrong citations based on the *Cited Reference Search mode* can also be included. The years of scientific life are taken to include the period from the first publication (this has been done here but sometimes the year in which the PhD was granted is taken) till present if the author is alive or till the year of his death (the latter applies to all authors in Table 3 except Ginzburg and Hänsch). The square in the denominator of the expression has been included to keep IA constant whenever the number of citations per publication and the number of publications per year remains constant throughout the scientific life.

The highest IA index in the upper part of Table 3 is that of Landau (6.19). This number is, however, rather low compared with present day authors of similar caliber (The IA index of this year's Nobel laureate T.W. Hänsch is 9.57). This effect is even more striking in the case of H. Bethe (3.46) who enjoyed a rather long (scientific) life, thus enhancing the denominator of the above expression while his productivity decreased in the later years of his life. Amaldi stayed in Italy during and after the war. His productivity decreased significantly during this period.[20] Even during the recovery after 1950, his productivity remained low. He was heavily involved in rebuilding Italian science at the political and administrative level, including the promotion of large national and international projects (CERN, Frascati, the European Space Agency). He was also involved in projects and workshops dealing with disarmament and the peaceful use of atomic energy, obviously at the expense of conventional scientific research. Hence his low IA number (0.40). The other IA giants in Table 3 are E. Teller (5.35) and L. Van Hove (5.90). The IA of G. Placzek (2.0) falls somewhere in the middle. It would be increased by about a factor of two if his highly cited book and book article were included.

# 7. Conclusions

We have performed a bibliometric (citations) analysis of Placzek's scientific publications. His variegated and tragic life offers an excellent case study for such analysis. We conclude that Placzek ranks according to standard bibliometric impact indexes, and in spite of his untimely death, among the giants of his life period, many of whom are found as coauthors in his publications.

except in very fragmentary form as a Letter to the Editor [G. Placzek, Phys. Rev. **93**, 895 (1954)]. In view of the importance of these results for actual computation of slow-neutron scattering cross sections, it was considered useful to publish them after Dr. Placzek's death, as a complement to the above-mentioned letter. The author's original notes have been reviewed and edited for publication by L. Van Hove, Utrecht, Netherlands."

[12] See, for instance, the Springer Series Light Scattering in Solids, Vols. 1-9, which was launched in 1975.

[13] A.F.J. van Raan, Scientometrics **59**, 461 (2004).

[14] Note that the ordinate of this figure corresponds to the number of citing papers which, as already mentioned, may be somewhat lower than the number of citations.

[15] Caveat: The abstracts are only stored in the WoS after 1990. A given name can appear as an informal and a formal citation simultaneously, although this case is not very common.

[16] J. Hirsch, Proc. Nat. Acad. Sciences **102**, 16569 (2005).

[17] O.K. Andersen, private communication.

[18] Generally, papers in the biomedical field are cited twice as much as those in either physics or chemistry.

[19] C.V. Raman, Nature **121**, 501 (1928).

[20] M. Cardona and W. Marx, Scientometrics **64**, 313 (2005).

Corresponding author:
Prof. Dr. Manuel Cardona
Max Planck Institute for Solid State Research
Heisenbergstrasse 1, D-70569 Stuttgart
E-mail address: m.cardona@fkf.mpg.de